\documentclass[10pt,twocolumn,letterpaper]{article}

\usepackage[pagenumbers]{wacv} 

\definecolor{wacvblue}{rgb}{0.21,0.49,0.74}
\usepackage[pagebackref,breaklinks,colorlinks,allcolors=wacvblue]{hyperref}

\def\wacvPaperID{*****} 
\def\confName{WACV}
\def\confYear{2027}

\title{Longitudinal Lesion Inpainting in Brain MRI via
3D Region Aware Diffusion}

\author{
Zahra Karimaghaloo\textsuperscript{1} \quad
Haz-Edine Assemlal\textsuperscript{1} \quad
Dumitru Fetco\textsuperscript{1} \quad
Hassan Rivaz\textsuperscript{2} \quad
Douglas L. Arnold\textsuperscript{1}\\
\textsuperscript{1}Clario part of Thermofisher Scientific, Montreal, Canada \quad
\textsuperscript{2}Concordia University, Montreal, Canada\\
{\tt\small \{zahra.karimaghaloo, haz-edine.assemlal, dumitru.fetco, douglas.arnold\}@clario.com}\\
{\tt\small hassan.rivaz@concordia.ca}
}

\begin{document}
\maketitle
\begin{abstract}

Accurate longitudinal analysis of brain MRI is often hindered by evolving lesions, which bias automated neuroimaging pipelines. While deep generative models have shown promise in inpainting these lesions, most existing methods operate cross-sectionally or lack 3D anatomical continuity. We present a novel pseudo-3D longitudinal inpainting framework based on Denoising Diffusion Probabilistic Models (DDPM). Our approach utilizes multi-channel conditioning to incorporate longitudinal context from distinct visits ($t_1, t_2$) and extends Region-Aware Diffusion (RAD) to the medical domain, focusing the generative process on pathological regions without altering surrounding healthy tissue. We evaluated our model against state-of-the-art baselines on longitudinal brain MRI from 93 patients. Our model significantly outperforms the leading baseline (FastSurfer-LIT) in terms of perceptual fidelity, reducing the LPIPS distance from 0.07 to 0.03 while eliminating inter-slice discontinuities. Furthermore, our framework demonstrates high longitudinal stability with a Temporal Fidelity Index of 1.02, closely approaching the ideal value of 1.0 and substantially narrowing the gap compared to LIT's TFI of 1.22. Our proposed model also achieves an average processing time of 2.53 min per volume, representing approximately 10$\times$ speedup over LIT. Crucially, we validated the downstream impact of our method on volume change computation and non-linear registration. 
As a key downstream validation, we track geometric distortion via the log-Jacobian determinant across multiple concentric peri-lesional rings. Our results show that the proposed paradigm significantly reduces downstream registration artifacts over baselines, preserving structural fidelity across both central pathology and surrounding tissue boundaries.
By leveraging longitudinal priors and region-specific denoising, our framework provides a highly reliable and efficient preprocessing step for studying progressive neurodegenerative diseases. A derivative dataset of 93 scans and our code will be available upon acceptance.

\end{abstract}
    
\section{Introduction}
\label{sec:intro}

The longitudinal analysis of structural brain MRI is essential for monitoring neurodegenerative diseases. However, the appearance of lesions, whether new, resolving, or changing in size and/or pattern, often biases automated pipelines, such as longitudinal registration and atrophy measurements. To mitigate this, image inpainting, also known as lesion filling, replaces pathological regions with healthy-appearing tissue. 

Widely-used neuroimaging toolboxes such as NiftySeg \cite{cardoso2015} perform subject-specific optimization for joint tissue segmentation and lesion filling, often leveraging non-local means strategies \cite{guizard2015non}. While effective for small lesions, these methods struggle with complex anatomical structures or large cavities where local texture similarity is insufficient.

Recent work has shifted toward deep generative models, with Denoising Diffusion Probabilistic Models (DDPMs) setting new benchmarks~\cite{lei2025lesiondiffusion,seo2025diffusion,liu2025pathopainter, ho2020denoising, zhang2026hieredit, liu2026region, vu2026inverfill}. RePaint \cite{lugmayr2022repaint} demonstrated that unconditional diffusion priors could be repurposed for inpainting, while BrushNet~\cite{ju2024brushnet} introduced dual-branch architectures to separate masked image features from noisy latents for cleaner preservation of unmasked regions. In the medical domain, FastSurfer-LIT \cite{pollak2024} introduced a resolution-independent DDPM for brain lesion inpainting, and LG-Net~\cite{tang2021lg} utilized learnable dynamic gate convolutions with feature-consistency losses specifically tailored for Multiple Sclerosis (MS) pathology. FastSurfer-LIT serves as the primary modern diffusion baseline for our evaluation, as it represents the leading resolution-independent generative approach in this domain. 

More recently, MSRepaint \cite{zhang2025arxive,zhang2025spie} have refined this using Conditional Denoising Diffusion Implicit Models (DDIMs). While these methods significantly improve anatomical fidelity, they are primarily cross-sectional or rely on multi-view averaging, which does not inherently enforce 3D volumetric continuity or utilize longitudinal priors. 

Furthermore, while longitudinal conditional DDPMs (CDDPMs) were previously introduced using 2D slice-based architectures \cite{zahra2026spie}, such methods lack the volumetric consistency required for precise medical analysis. Crucially, the downstream geometric impact of these inpainting techniques on non-linear image registration remains poorly validated. Left uncorrected, focal pathology distorts local deformation fields, introducing significant longitudinal bias that propagates far beyond the lesion core into surrounding healthy tissues. 

To address these limitations, we present a novel Pseudo-3D framework for longitudinal lesion inpainting in the context of Multiple Sclerosis (MS) disease. We utilize a multi-channel conditioning strategy that incorporates dual-timepoint ($t_1, t_2$) inputs and Cerebrospinal Fluid (CSF) masks for topological guidance. To optimize performance, we adapt the Region-Aware Diffusion (RAD) mechanism \cite{kim2025rad, liu2026region} to the medical domain. By applying a spatially variant noise schedule, our model focuses the reverse diffusion process strictly on pathological regions, ensuring surrounding healthy tissue remains unaltered.

We comprehensively evaluate the proposed framework by assessing both localized image quality and compatibility with downstream pipelines. Beyond establishing baseline performance through cross-sectional inpainting validation, we assess temporal consistency using an LPIPS-based longitudinal change preservation metric. Crucially, we extend this validation to downstream applications by tracking geometric distortion via the log-Jacobian determinant across multiple concentric peri-lesional rings. This structural analysis demonstrates that our paradigm significantly reduces downstream registration artifacts compared to baseline intensities and existing models, successfully maintaining anatomical fidelity across both central pathology and far-field tissue boundaries.

Our primary contributions are:
\begin{itemize}
    \item \textbf{Region-Aware Inpainting:} Adaptation of the RAD mechanism to 3D volumes, enabling spatially-variant denoising that limits updates to the lesion mask, significantly improving inference efficiency.
    \item \textbf{Joint Temporal Inpainting:} A simultaneous dual-timepoint processing framework that ensures inter-temporal consistency, mitigating longitudinal bias in downstream clinical measurements.
    \item \textbf{Dataset release:} To support benchmarking future studies, a task-specific derivative dataset consisting of pre-processed scans from longitudinal MRI scans of 93 patients used for testing is available upon request, subject to approval by the study sponsors and in accordance with the anonymous data sharing policies.
\end{itemize}

\section{Methods}
\label{sec:Methods}
\begin{figure*}[!htb]
\centering
    \includegraphics[width=.7\linewidth]{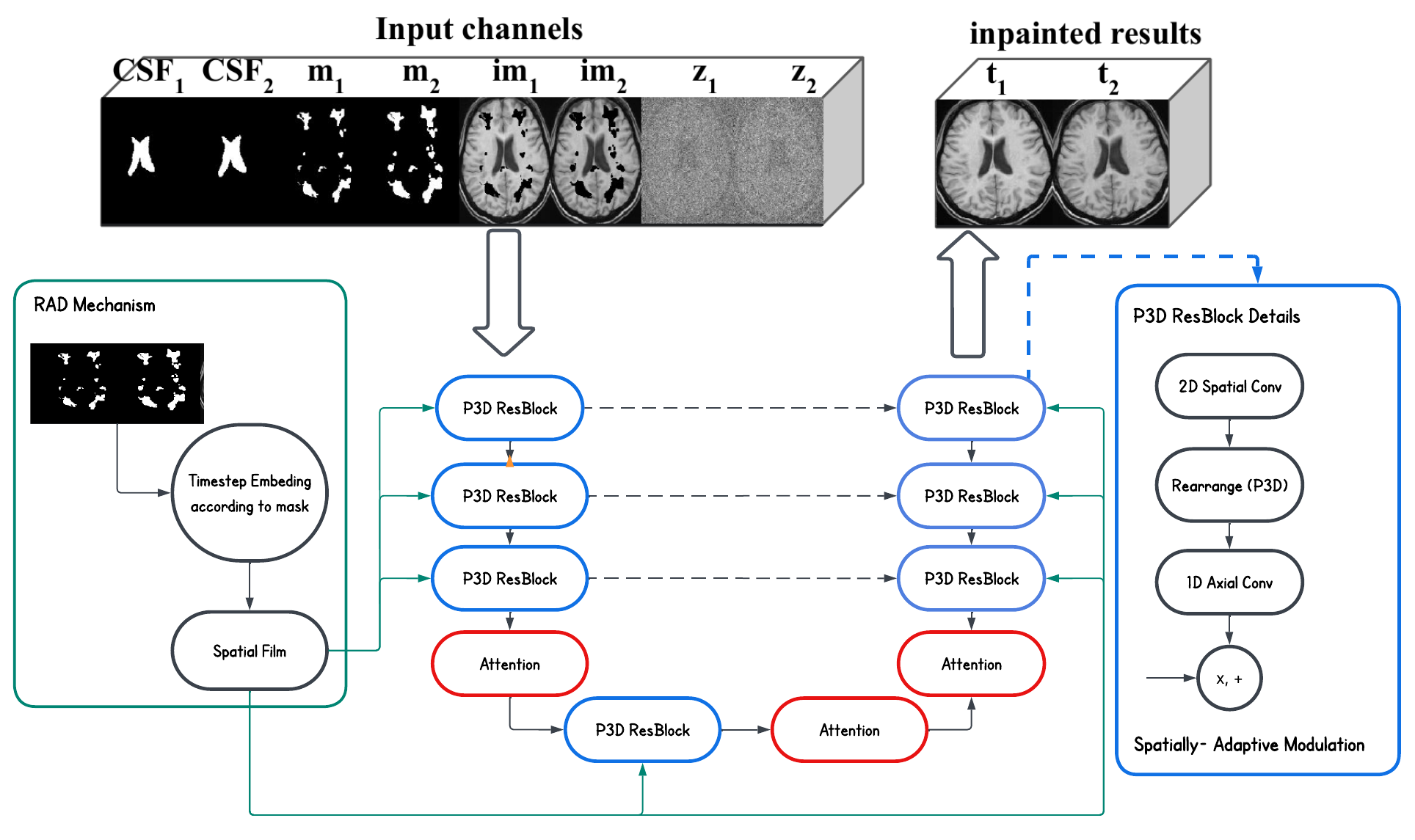}
\caption{P3D-RAD Architecture. The model integrates Pseudo-3D convolutions with a Region Aware Diffusion (RAD) mechanism. Spatial FiLM parameters are injected into both the encoder and decoder P3D ResBlocks (green lines) to provide temporally and spatially aware feature modulation.}
    \label{fig:flowchart}

\end{figure*}



Advancing beyond traditional 2D multi-channel methods, we introduce a Pseudo-3D Conditional DDPM (P3D-CDDPM) that leverages inter-slice context for longitudinal lesion inpainting by simultaneously processing two temporal volumes ($t_1, t_2$). We further extend this framework into P3D-RAD, incorporating a Region-Aware Denoising (RAD) mechanism to optimize generative efficiency and enhance anatomical fidelity.
Following~\cite{zahra2026spie}, the input tensor $\mathbf{X} \in \mathbb{R}^{8 \times D \times H \times W}$ provides longitudinal context by concatenating eight channels: two CSF anatomical priors, two binary lesion masks $\mathbf{m}_{\{1,2\}}$, two masked images $\mathbf{im}_{\{1,2\}}$ (lesions zero-filled), and two diffusion noise components $\mathbf{z}_{\{1,2\}}$ (Fig.~\ref{fig:flowchart}).

\subsection{Conditional Diffusion Preliminaries.}
A denoising diffusion probabilistic model (DDPM)~\cite{ho2020denoising} learns a data distribution $q(\mathbf{x}_0)$ by reversing a fixed Markovian forward process that gradually corrupts a clean image $\mathbf{x}_0$ into Gaussian noise. The forward process is defined as
\begin{equation}
\begin{aligned}
    q(\mathbf{x}_{1:T} \mid \mathbf{x}_0) &= \prod_{t=1}^{T} q(\mathbf{x}_t \mid \mathbf{x}_{t-1}), \\
    q(\mathbf{x}_t \mid \mathbf{x}_{t-1}) &= \mathcal{N}\!\left(\sqrt{1-\beta_t}\,\mathbf{x}_{t-1},\, \beta_t \mathbf{I}\right),
\end{aligned}
\end{equation}
which results in the closed-form marginal $q(\mathbf{x}_t \mid \mathbf{x}_0) = \mathcal{N}(\sqrt{\bar{\alpha}_t}\,\mathbf{x}_0,\, (1-\bar{\alpha}_t)\mathbf{I})$, with $\alpha_t \triangleq 1-\beta_t$ and $\bar{\alpha}_t \triangleq \prod_{s=1}^{t}\alpha_s$. The reverse process is parameterized as a learned Gaussian Markov chain, and the network $\boldsymbol{\epsilon}_\theta$ is trained to predict the injected noise. In our setting, generation is conditioned on the longitudinal context $\mathbf{X}_{\text{cond}}$ (the CSF priors, lesion masks, and masked images), which is supplied to the network by channel concatenation. The training objective is thus
\begin{equation}
    \mathcal{L} = \mathbb{E}_{\mathbf{x}_0, \boldsymbol{\epsilon}, t}\left[\, \big\| \boldsymbol{\epsilon} - \boldsymbol{\epsilon}_\theta(\mathbf{x}_t,\, t \mid \mathbf{X}_{\text{cond}}) \big\|^2 \,\right], \quad \boldsymbol{\epsilon} \sim \mathcal{N}(\mathbf{0}, \mathbf{I}),
\end{equation}
where $\mathbf{x}_t = \sqrt{\bar{\alpha}_t}\,\mathbf{x}_0 + \sqrt{1-\bar{\alpha}_t}\,\boldsymbol{\epsilon}$. At inference, the masked tissue is recovered by iterating the learned reverse transitions from $t=T$ down to $t=0$ while keeping $\mathbf{X}_{\text{cond}}$ fixed.

\subsection{Pseudo-3D Architecture.}
Leveraging a pseudo-3D design as described in \cite{durrer2024denoising,zhu2023make}, our model captures essential inter-slice context through 1D axial convolutions, providing a computationally efficient alternative to fully 3D U-Net implementations. We feed a stack of adjacent axial slices into the network, allowing the model to recognize 3D anatomical structures while maintaining the efficiency of 2D generative kernels. We utilize a (2+1)D convolution strategy within each residual block to enforce inter-slice continuity. Formally, the feature transformation is defined as
\begin{equation}
    \text{P3D}(\mathbf{h}) = \text{Conv}_{1D}\!\left(\mathcal{R}\!\left(\text{Conv}_{2D}(\mathbf{h})\right)\right),
\end{equation}
where $\text{Conv}_{2D}$ operates on in-plane $(H,W)$ features, $\mathcal{R}$ denotes the dimension rearrangement required to expose the axial ($z$) dimension, and $\text{Conv}_{1D}$ aggregates context across slices. This factorization captures 3D anatomical context while maintaining the efficiency of 2D generative kernels, effectively resolving ``staircase'' artifacts.

To train this factorized kernel, we adopt a two-stage curriculum. We first learn a purely 2D slice-wise inpainting model, optimizing the in-plane $\text{Conv}_{2D}$ weights independently of any axial coupling. We then introduce the axial $\text{Conv}_{1D}$ branch along the $z$ dimension, initializing the network from the learned 2D weights and subsequently optimizing all parameters jointly. This warm-start strategy lets the model inherit strong in-plane priors before adapting them to enforce inter-slice continuity, stabilizing training of the volumetric kernel.

\subsection{Region Aware Diffusion (RAD).}
In our P3D-RAD model, we extend RAD~\cite{kim2025rad} into a Pseudo-3D (P3D) architecture to capture volumetric context for longitudinal MRI. Unlike vanilla diffusion, which applies an identical scalar noise schedule to every voxel, RAD assigns a spatially variant schedule so that distinct regions can be generated asynchronously while still attending to global context. Concretely, the per-element forward transition becomes
\begin{equation}
    q(\mathbf{x}_{t,i} \mid \mathbf{x}_{t-1,i}) = \mathcal{N}\!\left(\sqrt{1-b_{t,i}}\,\mathbf{x}_{t-1,i},\, b_{t,i}\right),
\end{equation}
with marginal $q(\mathbf{x}_{t,i} \mid \mathbf{x}_{0,i}) = \mathcal{N}(\sqrt{\bar{a}_{t,i}}\,\mathbf{x}_{0,i},\, 1-\bar{a}_{t,i})$, where $a_{t,i} \triangleq 1-b_{t,i}$, $\bar{a}_{t,i} \triangleq \prod_{s=1}^{t} a_{s,i}$, and $\bar{b}_{t,i} \triangleq 1-\bar{a}_{t,i}$ is the pixel-wise accumulated noise intensity. RAD realizes inpainting through a two-phase noise schedule: during training, noise is added to the inpainting (lesion) region in Phase 1 ($t \in [0, T/2]$) and to the surrounding background in Phase 2 ($t \in [T/2+1, T]$). At inference, we bypass Phase 2 and initialize the reverse process at $t=T/2$, ensuring the healthy background remains untouched while generative capacity is spent solely on the masked tissue.

To inform the denoiser of these voxel-specific noise levels, RAD replaces the scalar timestep embedding with the pixel-wise intensity map $\bar{b}_t$. In our P3D-RAD instantiation, this regional conditioning is integrated into the P3D ResBlock via Spatial FiLM~\cite{perez2018film} modulation (Fig.~\ref{fig:flowchart}). Following the 1D axial convolution, the spatial timestep map $t(s)$, derived from the local accumulated noise $\bar{b}_t$, is injected to modulate the feature maps as
\begin{equation}
    \mathbf{h}_{out}(s) = \gamma(t(s)) \odot \mathbf{h}_{in}(s) + \beta(t(s)),
\end{equation}
where $\gamma$ and $\beta$ are scaling and shifting parameters learned from the local noise levels. This makes the (2+1)D kernels aware of inpainting boundaries, focusing generative capacity on the intended regions.

\section{Experiments and Results}
\label{sec:experiments}

To evaluate the effectiveness of our proposed framework, we compare several architectural configurations and inference strategies, as outlined below. The MRI sequence used for all models consisted of T1-weighted (T1w) images from both timepoints, which were linearly co-registered to a common space to maintain anatomical consistency. Ground truth data were generated by transplanting lesion masks from an external cohort onto the test scans. This synthesis was constrained to the brain mask, ensuring that artificial lesions were placed exclusively in valid tissue regions while avoiding cerebrospinal fluid (CSF) and existing pathology.

\begin{itemize}
    \vspace{2mm}
    \item \textbf{CDDPM vs. RAD:} CDDPM serves as the baseline global denoising strategy, wherein the network reconstructs the entire volume; the generated content within the inpainting mask is then fused with the original healthy tissue during inference. In contrast, RAD utilizes spatially variant noise schedules to focus reverse steps strictly on the lesion mask. Both models were trained on 3D brain MRIs from 1,840 MS subjects (each with 2 timepoints) across varying disease stages and atrophy levels. 
    \vspace{2mm}

    \item \textbf{2D vs. P3D Architecture:} In the 2D configurations, models are trained on selected axial slices; at inference, the full 3D volume is reconstructed by sequentially processing every slice containing a mask. In contrast, pseudo-3D (P3D) models utilize 1D temporal convolutions to capture inter-slice continuity across the entire volume simultaneously.
    \vspace{2mm}

    \item \textbf{Other Baselines:} We compare our framework against several state-of-the-art baselines: LaMa~\cite{suvorov2022}, FastSurfer-LIT~\cite{pollak2024}, and RePaint~\cite{lugmayr2022repaint}. While LaMa (2D) and LIT (3D) were implemented using their respective official inference pipelines\footnote{Available at \url{https://github.com/advimman/lama} and \url{https://github.com/Deep-MI/FastSurfer}, respectively.}, RePaint required training from scratch to capture specific neuroanatomical priors for healthy T1-weighted brain tissue. We adopted the guided-diffusion U-Net architecture~\cite{dhariwal2021diffusion} and trained it using a large longitudinal MRI dataset of 1,840 MS subjects, using lesion-free slices and two time points per image. To adapt this cross-sectional architecture for longitudinal inpainting, we modified the input layer to a three-channel configuration comprising the T1w images at $t_1$ and $t_2$ along with their temporal difference ($t_2 - t_1$).
\end{itemize}

\begin{table*}[t]
\centering
\caption{Aggregate Quantitative Results: comparison of inpainting performance averaged across both $t_1$ and $t_2$.}
\label{tab:aggregate_results}
\resizebox{\textwidth}{!}{
\begin{tabular}{lccccccc}
\hline
\textbf{Model} & \textbf{NRMSE} $\downarrow$ & \textbf{PSNR} $\uparrow$ & \textbf{SSIM} $\uparrow$ & \textbf{LPIPS (Avg)} $\downarrow$ & \textbf{LPIPS (Axial)} & \textbf{LPIPS (Sagittal)} & \textbf{Time (min)} $\downarrow$ \\ \hline
\textit{Baselines} & & & & & & & \\
2D-RePaint & 0.108 $\pm$ 0.016 & 19.43 $\pm$ 1.40 & 0.60 $\pm$ 0.06 & 0.23 $\pm$ 0.05 & 0.18 $\pm$ 0.06 & 0.27 $\pm$ 0.05 & $> 2~hrs$ \\
2D-LaMa & 0.061 $\pm$ 0.010 & 24.47 $\pm$ 1.38 & 0.83 $\pm$ 0.04 & 0.10 $\pm$ 0.04 & 0.09 $\pm$ 0.04 & 0.12 $\pm$ 0.04 & 6.21 $\pm$ 0.12 \\
3D-LIT & 0.043 $\pm$ 0.011 & 27.49 $\pm$ 2.07 & 0.89 $\pm$ 0.04 & 0.07 $\pm$ 0.03 & 0.07 $\pm$ 0.03 & 0.07 $\pm$ 0.03 & 24.30 $\pm$ 1.84 \\ \hline
\textit{Ablations} & & & & & & & \\
2D-CDDPM & 0.044 $\pm$ 0.007 & 27.15 $\pm$ 1.43 & 0.86 $\pm$ 0.03 & 0.12 $\pm$ 0.04 & 0.08 $\pm$ 0.03 & 0.17 $\pm$ 0.05 & 7.77 $\pm$ 0.56 \\
2D-RAD & 0.050 $\pm$ 0.009 & 26.26 $\pm$ 1.66 & 0.86 $\pm$ 0.04 & 0.10 $\pm$ 0.03 & 0.06 $\pm$ 0.03 & 0.14 $\pm$ 0.05 & 4.42 $\pm$ 0.27 \\
P3D-CDDPM & 0.028 $\pm$ 0.004 & 31.18 $\pm$ 1.32 & 0.93 $\pm$ 0.02 & 0.04 $\pm$ 0.02 & 0.04 $\pm$ 0.02 & 0.04 $\pm$ 0.02 & 17.67 $\pm$ 0.75 \\
\textbf{P3D-RAD (Ours)} & \textbf{0.023 $\pm$ 0.004} & \textbf{32.82 $\pm$ 1.48} & \textbf{0.96 $\pm$ 0.01} & \textbf{0.03 $\pm$ 0.02} & \textbf{0.03 $\pm$ 0.02} & \textbf{0.03 $\pm$ 0.01} & \textbf{2.53 $\pm$ 0.70} \\ \hline
\end{tabular}
}
\end{table*}

\subsection{Cross-Sectional Anatomical Precision}
\begin{figure*}[tbp]
    \centering
    \includegraphics[width=.6\linewidth]{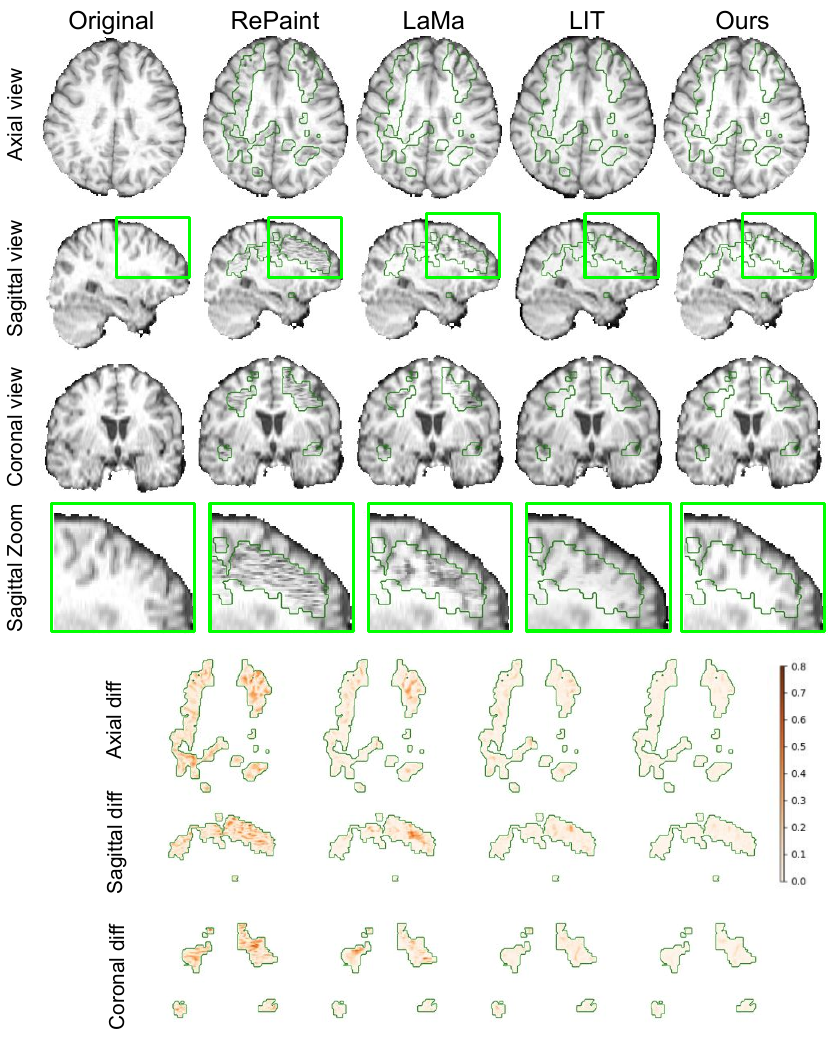}
    \caption{Qualitative comparison of 3D inpainting. Column 1 shows the ground truth; subsequent columns show model results (Rows 1–3) and their absolute error maps (Rows 5–7). Row 4 displays zoomed-in versions of the regions marked with light green within the sagittal view (Row 2). Dark green outlines the inpainting mask. As observed, our inpainting results show the highest fidelity to the ground truth.}
    \label{fig:xsec_visual_comp_dif}
\end{figure*}

We first assessed cross-sectional performance, measuring how closely inpainted regions at each visit ($t_1$ and $t_2$) resemble the known healthy anatomy of the ground truth. The longitudinal models (P3D-RAD, P3D-CDDPM, 2D-RAD, 2D-CDDPM, and RePaint) processed both timepoints simultaneously, while the cross-sectional baselines (LaMa and LIT) processed them independently.
Quantitative results are summarized in Table~\ref{tab:aggregate_results}, reporting mean performance across 93 unique subjects (each with 2 timepoints) over varying disease stages and atrophy levels. We use voxel-wise metrics, PSNR and NRMSE, to assess structural reconstruction, and perceptual metrics, SSIM and LPIPS, to evaluate texture realism. While FID is common for generative models, it is less suited here, as it assesses distribution-level rather than image-specific fidelity~\cite{adamson2025using}. LPIPS is a superior alternative for medical imaging, correlating more strongly with human perceptual judgment~\cite{konz2024rethinking,armanious2020medgan}. Since it operates on 2D patches, we compute it for inpainted slices across anatomical views and report the mean per volume.
Comparing 2D and Pseudo-3D (P3D) variants shows that 1D temporal convolutions substantially enhance spatial consistency. This is clearest in the sagittal view, where P3D-CDDPM reduced LPIPS from 0.17 to 0.04 over its 2D counterpart, and P3D-RAD from 0.14 to 0.03. The RAD mechanism also yields large efficiency gains, cutting inference time for two timepoints from 17.67 minutes (P3D-CDDPM) to 2.53 minutes (P3D-RAD).
Figure~\ref{fig:xsec_visual_comp_dif} compares P3D-RAD with the baselines qualitatively. Our model shows the highest anatomical fidelity, with minimal structural deviation from the ground truth, consistent with our expert rater evaluation, in which the vast majority of inpainted voxels were indistinguishable from real anatomy.

\begin{figure*}[tb]
    \centering
    \includegraphics[width=.6\linewidth]{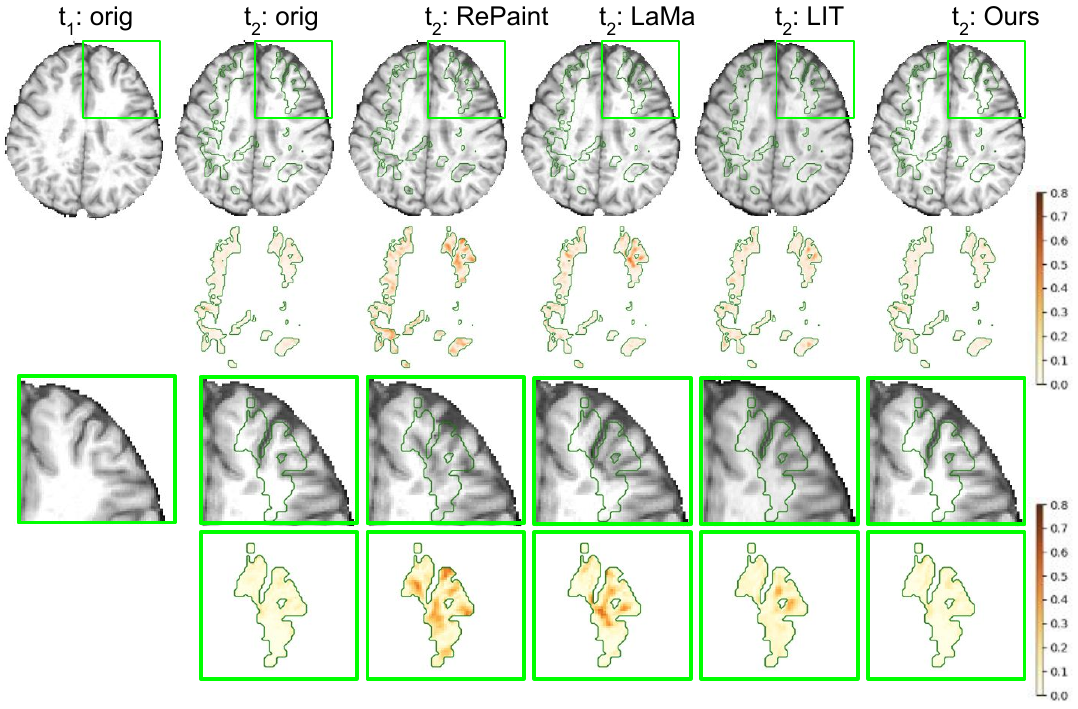}
    \caption{Longitudinal Inpainting. Inpainted $t_2$ across four models. Row 1: original at $t_1$ with no inpainting (same for all); Row 2: original and inpainted at $t_2$; Row 3: Absolute difference maps. Dark green outlines the inpainting mask. LIT appears blurred due to global blurring in its pipeline.}
    \label{fig:temporal_visual_comp}
\end{figure*} 

\subsection{Longitudinal Consistency and Efficiency}
\begin{figure*}[tb]
    \centering
    \includegraphics[width=.8\linewidth]{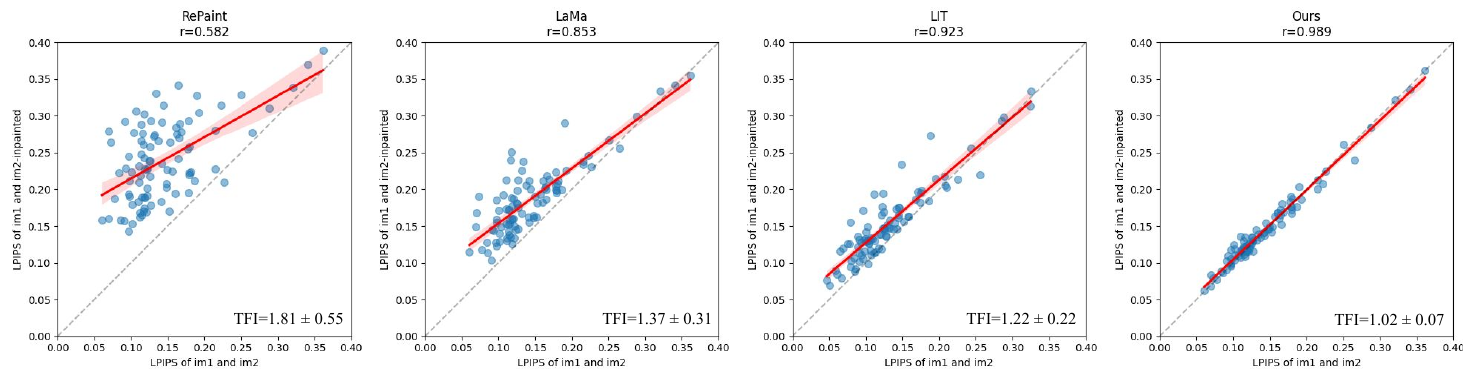}
    \caption{LPIPS-based longitudinal change preservation. Scatter compares original vs inpainted LPIPS between $(t_1,t_2)$. Red: linear fit; dashed black: identity ($y=x$). Our method achieves the highest Pearson ($r$) and Temporal Fidelity Index closest to 1.}
    \label{fig:temporal_lpips_ratio}
\end{figure*}

Beyond cross-sectional validation, we evaluate longitudinal consistency by assessing how models preserve patient-specific changes across visits. Fig.~\ref{fig:temporal_visual_comp} shows that while baselines fail to capture true temporal dynamics even in the axial plane, our model maintains high consistency with observed changes. In Fig.~\ref{fig:temporal_lpips_ratio}, we quantify this by correlating the LPIPS of inpainted pairs against that of the original pairs. Our framework achieves a Temporal Fidelity Index (TFI)---the ratio of these two LPIPS values---closest to the ideal 1.0, demonstrating superior preservation of disease progression with minimal artificial variance.

\begin{figure*}
    \centering
    \includegraphics[width=1\linewidth]{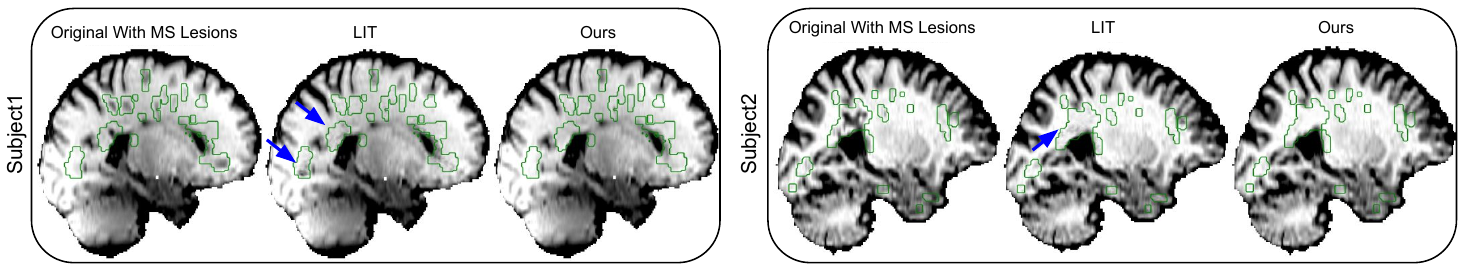}
    \caption{Expert Blind Review. Two examples of real MS lesion inpainting (green outlines). Blue arrows highlight structural artifacts in the LIT baseline identified by the expert.}
    \label{fig:blind_review_visual}
\end{figure*}

\subsection{Expert Blind Review}
To assess clinical realism, a blinded MRI expert rater, a physician evaluated 38 inpainted volumes from 14 MS subjects ($2\times$ timepoints). The evaluation was split into two cohorts: (1) synthetic lesions ($n=18$), where masks (with mean of $\sim$71k voxels; $\sim$16\% of white matter) were projected onto healthy tissue to test P3D-RAD's inpainting; and (2) real MS lesions ($n=10$), where actual pathology (with mean of $\sim$51k voxels; $\sim$12\% of white matter) was inpainted by both P3D-RAD and the LIT baseline for benchmarking. 

As shown in Table~\ref{tab:expert_blind_review}, the reader could not distinguish P3D-RAD's output from healthy anatomy, yielding negligible Dice scores for both synthetic ($0.001$) and real ($0.009$) cases. In contrast, LIT produced a significantly higher Dice score ($0.28$), as the expert identified structural inconsistencies, primarily in the sagittal view (Fig.~\ref{fig:blind_review_visual}). These results demonstrate that P3D-RAD produces anatomical textures indistinguishable from healthy tissue.

\begin{table}[tbp]
\centering
\caption{Expert Blind Review Results. Dice overlap between the expert's annotation of suspected inpainted tissue and the true inpainting mask. Lower scores are better and indicate that the expert could not localize the synthesized region. Volumes are in voxels.}
\label{tab:expert_blind_review}
\footnotesize
\setlength{\tabcolsep}{6pt}
\begin{tabular}{llcc}
\toprule
\textbf{Data Subset} & \textbf{Model} & \textbf{Overlap Volume} & \textbf{Dice Score $\downarrow$} \\ 
\midrule
\textbf{Synthetic} ($n=18$) & P3D-RAD & $30 \pm 22$ & $\mathbf{0.001 \pm 0.001}$ \\
\textit{Mask Vol. $\sim$71k} & & & \\
\midrule
\textbf{Real MS} ($n=10$) & P3D-RAD & $191 \pm 245$ & $\mathbf{0.008 \pm 0.011}$ \\
\textit{Mask Vol. $\sim$51k} & LIT & $9,501 \pm 3,313$ & $0.282 \pm 0.091$ \\ 
\bottomrule
\end{tabular}
\end{table}

\subsection{Non-Linear Registration and Jacobian Validation Framework}
\begin{figure*}[tb]
    \centering
    \includegraphics[width=1\linewidth]{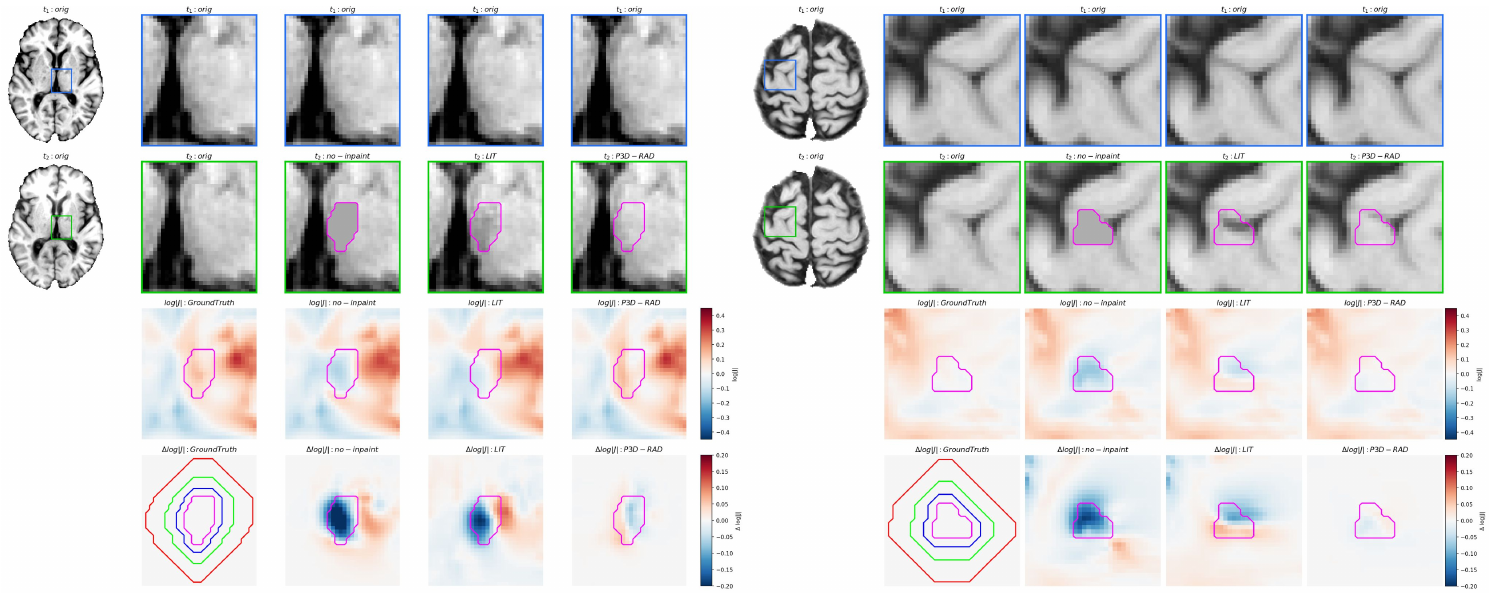}
    \caption{Qualitative comparison of inpainting effects on the downstream task of Jacobian ($J(\mathbf{x})$) computation across two examples.
    For each example, the first column displays the full original structural images. The subsequent columns show close-up views corresponding to the marked rectangular regions where the lesion is inserted. 
    Within these close-ups, the first row displays the original image at $t_1$, while the second row shows the original image at $t_2$ alongside the lesion-inserted (no-inpaint), LIT, and P3D-RAD inpainted views. 
    The third row displays the $\log|J(\mathbf{x})|$ maps, and the final row illustrates the deviation ($\Delta\log|J(\mathbf{x})|$) of each approach from the ground truth (where values closer to 0 are ideal). Concentric ROI rings centered around the lesion area provide a spatial reference for evaluation. The area marked in magenta is the core lesion.}
    \label{fig:log_jac_example}
\end{figure*}

To evaluate the downstream clinical utility of lesion inpainting in mitigating artificial geometric distortion during non-linear image registration, we established a longitudinal validation framework. The objective was to quantify how temporary focal pathology biases local deformation fields, measured via the log-Jacobian determinant, and to demonstrate our model's structural restoration performance.
Rather than using synthetic geometric shapes, realistic focal lesion masks were derived from an independent cohort of real MS patient scans, ensuring that the morphology, volumes, and shapes of evaluated pathologies faithfully represent true clinical presentation. A selective insertion protocol was enforced: lesions were transplanted exclusively into lesion-free white matter and juxtacortical regions of the host images, which have high clinical susceptibility for MS pathology.
To capture spatial evolution dynamics common in neuroimaging, two longitudinal scenarios were simulated ($N = 59$ total cases):
\textbf{New Lesion ($n = 28$):} acute development, where pathology is present exclusively at the second timepoint ($\text{t}_2$).
\textbf{Resolving Lesion ($n = 31$):} resolution, where the lesion is present solely at baseline ($\text{t}_1$).
For each case, four variants of longitudinal image pairs were generated:
\begin{itemize}
    \item \textbf{Ground-Truth:} The original, unaltered scans, establishing baseline volume changes within the selected regions of interest (ROIs).
    \item \textbf{Lesion-Affected (No-Inpaint):} Uncorrected volumes where the lesion mask is filled with mean gray matter intensity at the relevant visit (New or Resolving).
    \item \textbf{LIT Inpainted:} Volumes corrected with the baseline LIT algorithm prior to registration.
    \item \textbf{P3D-RAD Inpainted:} Volumes corrected with our proposed framework prior to registration.
\end{itemize}
Since all pairs were already linearly aligned, non-linear registration was performed via Symmetric Normalization (SyN) using ANTsPy~\cite{tustisonantsx}, with $t_2$ as the fixed reference and $t_1$ as the moving image.
Registration yields a continuous deformation field, $\phi: \mathbb{R}^3 \to \mathbb{R}^3$, mapping the moving image to the fixed reference. Local distortions are captured by the Jacobian $J(\mathbf{x}) = \nabla\phi(\mathbf{x})$, and the log-Jacobian determinant $\log|J(\mathbf{x})|$ quantifies voxel-wise volume changes, where $\log|J(\mathbf{x})| > 0$ denotes expansion and $\log|J(\mathbf{x})| < 0$ contraction.
Figure~\ref{fig:log_jac_example} shows two qualitative examples. For each subject, the $\log|J(\mathbf{x})|$ map from the original pair serves as ground truth. As seen in the third row, leaving the lesion un-inpainted introduces severe distortion. LIT inpainting markedly reduces this error, while our P3D-RAD paradigm yields results highly consistent with the ground-truth Jacobian, as illustrated in the final row mapping each method's voxel-wise difference of $\log|J(\mathbf{x})|$ relative to the ground truth.

To systematically track how registration artifacts propagate from the core focal pathology into surrounding healthy structural tissue, concentric ROIs were defined relative to the transplanted lesion boundary:
\begin{itemize}
    \item \textbf{Core:} The localized mask representing the exact transplanted lesion volume.
    \item \textbf{Ring 1, 2:} The immediate peri-lesional boundary, defined by a 1- and 2-voxel morphological dilation of the core, respectively, excluding the core mask itself.
    \item \textbf{Ring 5, 10:} Wide-radius peripheral bands defined by 5- and 10-voxel dilations, respectively, designed to capture long-range or far-field deformation field distortions.
    \item \textbf{Core + Ring 5 / Ring 10:} Aggregate global regions evaluating the compounded impact across both the central lesion and its surrounding anatomy.
\end{itemize}
These ROIs are marked in Figure~\ref{fig:log_jac_example}. For each ROI, we computed the Mean Absolute Difference (MAD) of $\log|J(\mathbf{x})|$ relative to the original deformation, defined as:
\begin{equation}
    \text{MAD}_m = \frac{1}{N_{\text{ROI}}} \sum_{\mathbf{x} \in \text{ROI}} \left| \log|J_m(\mathbf{x})| - \log|J_{\text{orig}}(\mathbf{x})| \right|.
\end{equation}
Table~\ref{tab:mad_per_region} summarizes the mean and standard deviation of the MAD measurement for all 59 cases across the core, all concentric rings, and the aggregate regions. 


Our proposed method yields the lowest MAD across every evaluated region, indicating that it perturbs the underlying deformation field substantially less than competing baselines. The effect is most pronounced in the lesion \emph{core}, where P3D-RAD reduces the MAD by roughly $38\%$ relative to LIT ($0.019$ vs.\ $0.031$) and by nearly an order of magnitude relative to the uncorrected baseline ($0.019$ vs.\ $0.130$). This competitive advantage persists robustly across all concentric peripheral rings. Paired Wilcoxon signed-rank tests confirm that P3D-RAD errors are significantly lower than both LIT and no-inpaint configurations across all regions ($p < 0.001$ for all comparisons following Holm correction).

\begin{table*}[t]
\centering
\caption{Per-region Mean Absolute Difference (MAD) in $\log|J(\mathbf{x})|$ relative to the original baseline deformation, averaged over 59 evaluation cases ($\text{mean} \pm \text{std}$). The proposed P3D-RAD approach is evaluated against the uncorrected (no-inpaint) and the baseline LIT models. Lower values indicate better performance; the top-performing approach per row is highlighted in \textbf{bold}. For every evaluated region, P3D-RAD achieves a significantly lower error rate than both baselines ($p < 0.001$, paired Wilcoxon signed-rank test with Holm correction).}
\label{tab:mad_per_region}
\begin{tabular}{lccc}
\toprule
Region & MAD$_{\text{no-inpaint}}$ & MAD$_{\text{LIT}}$ & MAD$_{\text{P3D-RAD}}$ \\
\midrule
Core               & 0.130 $\pm$ 0.052 & 0.031 $\pm$ 0.017 & \textbf{0.019 $\pm$ 0.011} \\
Ring 1             & 0.085 $\pm$ 0.034 & 0.022 $\pm$ 0.012 & \textbf{0.013 $\pm$ 0.008} \\
Ring 2             & 0.072 $\pm$ 0.029 & 0.019 $\pm$ 0.010 & \textbf{0.011 $\pm$ 0.007} \\
Ring 5             & 0.048 $\pm$ 0.017 & 0.014 $\pm$ 0.006 & \textbf{0.007 $\pm$ 0.004} \\
Ring 10            & 0.029 $\pm$ 0.009 & 0.010 $\pm$ 0.004 & \textbf{0.004 $\pm$ 0.002} \\
Core + Ring 5      & 0.060 $\pm$ 0.021 & 0.016 $\pm$ 0.008 & \textbf{0.009 $\pm$ 0.005} \\
Core + Ring 10     & 0.034 $\pm$ 0.011 & 0.012 $\pm$ 0.004 & \textbf{0.005 $\pm$ 0.003} \\
\bottomrule
\end{tabular}
\end{table*}

\subsection{Implementation Details}
The P3D-RAD model was trained on an NVIDIA H100 multi-GPU cluster using the Adam optimizer with a learning rate of $10^{-4}$ for 300 epochs. To capture volumetric context, we employed a window size of 32, processing consecutive axial slices during training. At inference, the model processed the full stack of masked slices on a single GPU. We utilized the velocity ($v$-prediction)~\cite{karras2022elucidating} objective for the diffusion process, optimized via the Min-SNR weighting strategy ($\gamma=5$) to balance training stability across timesteps~\cite{hang2023efficient}. While training was conducted over $T=1000$ steps, our experiments demonstrated that 100 reverse steps were sufficient for P3D-RAD for high-fidelity inpainting at inference. CSF masks were generated using SynSeg-Net~\cite{huo2018synseg}, and lesion masks via in-house automatic detection followed by expert manual correction.
\section{Conclusions}

We presented P3D-RAD, a longitudinal inpainting framework that combines 1D axial convolutions with a Region Aware Diffusion mechanism. The model captures essential volumetric context and enforces anatomical continuity without the computational cost of native 3D U-Nets. Quantitatively, P3D-RAD significantly outperformed state-of-the-art baselines, reducing non-axial perceptual error (LPIPS) by nearly 80\% compared to 2D variants. Crucially, the downstream evaluation of voxel-wise log-Jacobian determinant maps demonstrates that P3D-RAD preserves underlying deformation fields significantly better than existing approaches, yielding the lowest Mean Absolute Difference (MAD) across all evaluated lesion cores and peripheral regions. By tightly bounding registration errors even within immediate peri-lesional bands, the joint inpainting of both time points ensures high longitudinal fidelity, preserving the patient's unique anatomical signature and preventing synthetic artifacts that could bias clinical atrophy measurements. By focusing the diffusion process on pathological regions, our framework eliminates post-hoc blending and minimizes inference time. This balance of spatial consistency and computational efficiency makes P3D-RAD a robust solution for integration into clinical pipelines for monitoring disease progression and automated brain volume quantification. Finally, we will release our test dataset (subject to approval by the study sponsors and in accordance with the anonymous data sharing policies) to benchmark future studies.

{
    \small
    \bibliographystyle{ieeenat_fullname}
    \bibliography{main}
}

\end{document}